# STATUS AND UPDATE OF THE RaDIATE COLLABORATION R&D PROGRAM

**K. Ammigan and P. Hurh**
Fermi National Accelerator Laboratory*
P.O Box 500, Batavia, IL 60510
ammikav@fnal.gov; hurh@fnal.gov

**ABSTRACT**

The Radiation Damage In Accelerator Target Environments (RaDIATE) collaboration was founded in 2012 and currently consists of over 50 participants and 11 institutions globally. Due to the increasing power of future proton accelerator sources in target facilities, there is a critical need to further understand the physical and thermo-mechanical radiation response of target facility materials. Thus, the primary objective of the RaDIATE collaboration is to draw on existing expertise in the nuclear materials and accelerator targets fields to generate new and useful materials data for application within the accelerator and fission/fusion communities. Current research activities of the collaboration include post irradiation examination (PIE) of decommissioned components from existing beamlines such as the NuMI beryllium beam window and graphite NT-02 target material. PIE of these components includes advanced microstructural analyses (SEM/TEM, EBSD, EDS) and micro-mechanics technique such as nano-indentation, to help characterize any microstructural radiation damage incurred during operation. New irradiation campaigns of various candidate materials at both low and high energy beam facilities are also being pursued. Beryllium helium implantation studies at the University of Surrey as well as high energy proton irradiation of various materials at Brookhaven National Laboratory's BLIP facility have been initiated. The program also extends to beam-induced thermal shock experiments using high intensity beam pulses at CERN's HiRadMat facility, followed by advanced PIE activities to evaluate thermal shock resistance of the materials. Preliminary results from ongoing research activities, as well as the future plans of the RaDIATE collaboration R&D program will be discussed.

1. INTRODUCTION

The Radiation Damage In Accelerator Target Environments (RaDIATE) collaboration was formed soon after the Proton Accelerators for Science and Innovation (PASI) workshop held at Fermilab in 2012. The PASI workshop participants, representing several high power accelerator facilities, identified radiation damage and thermal shock response as the most cross-cutting technical challenges facing accelerator components in future high power target facilities [1]. Subsequently, the RaDIATE collaboration was established by bringing together experts from both the nuclear materials and accelerator target facilities fields to address the relevant challenges and maximize the benefits of high power accelerators. Since its inception, the collaboration and its research activities have grown and now include 14 international participating institutions.

Radiation damage due to high energy particle-matter interactions occurs in the form of atomic displacements and creation of transmutation products, both of which disrupt the microstructure of the material and, in turn, affect critical bulk mechanical and thermal properties. Radiation damage is dependent on several irradiation parameters such as the irradiation temperature, particle flux (dose rate), and transmutation gas production. Target accelerator environments have relatively higher dose rate, higher gas production, and pulsed irradiation conditions than the lower dose rate, lower gas production and continuous irradiation present in the nuclear power environment. As a result, material properties of interest between the nuclear and accelerator applications are somewhat different.

Accelerator target and beam window are also subjected to highly localized, cyclic thermal gradients from the pulse nature of the irradiation. This results in what is referred to as thermal shock, where dynamic stress waves are generated and propagate through the material. Therefore, in the accelerator target environment, high-cycle fatigue and thermal diffusion are primary concerns, and accounting for the corresponding property degradation due to radiation damage is critical. The RaDIATE collaboration aims to identify and perform research activities tailored specifically to the high power accelerator target and beam window applications, relevant to several current and future accelerator facilities.

2. CURRENT RADIATE ACTIVITIES

Over the last few years, the RaDIATE collaboration has initiated research activities to address radiation damage and thermal shock in candidate materials for beam intercepting components (beam windows, secondary particle generation targets) at various facilities. These include conventional materials such as graphite, beryllium, aluminum, and titanium alloys, as well as novel potential target or window materials such as nano-fiber electro-spun mats and glassy carbon. The status and highlights of the ongoing studies are given in the following sections.

2.1. High-Energy Proton Irradiation of Graphite

High energy proton (181 MeV) irradiation of various graphite grades were carried out at the Brookhaven National Laboratory's Linac Isotope Producer facility (BLIP). The fine-grained, isotropic graphite grades irradiated included the POCO ZXF-5Q, Toyo-Tanso IG-430, Carbone-Lorraine 2020, and the SGL R7650, all candidate target materials for the Long Baseline Neutrino Facility (LBNF) [3]. The specimens were exposed to a peak fluence of $6.7 \times 10^{20}$ protons/cm$^2$ or about 0.1 DPA (displacements per atom) at an irradiation temperature of 120-150 °C. Figure 1 shows an open capsule containing tensile specimens being recovered after irradiation in the BLIP hot cell.

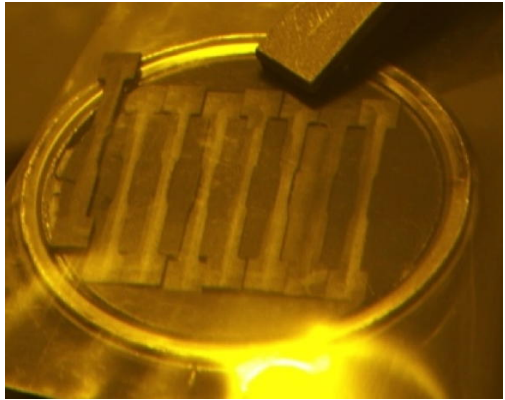

Figure 1. Graphite specimens being recovered after irradiation at BLIP (BNL).

After sufficient cool-down of the specimens, post-irradiation examination (PIE) of the different specimen grades were carried out [3,4]. The tensile strength and elastic modulus were seen to increase by about 30-50% after irradiation up to 0.056 DPA at 125 – 175 ˚C. Annealing above the irradiation temperature demonstrated partial recovery of initial properties, as shown in Figure 2.

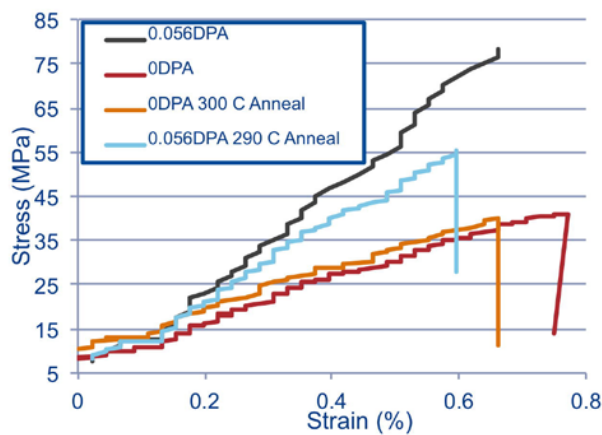

Figure 2. Tensile behavior of graphite specimens before and after irradiation at BLIP.

Measurement of the coefficient of thermal expansion revealed a 5-20% increase in value depending on the level of DPA accumulated in the specimens. Energy-dispersive x-ray diffraction (EDXRD) examination at BNL's NSLS-1 facility was also carried out. Results indicated about 3% lattice swelling, in agreement with neutron irradiated graphite data. Detailed results and discussions of PIE work on the various irradiated graphite grades are given in [4]. Future work will include a second irradiation experiment carried out at elevated temperature (up to 1000 °C) and the recreation of the damaged microstructure using low energy ion irradiation methods.

## 2.2. NT-02 NuMI Graphite Target Fin Study

The NT-02 neutrino production target in the NuMI beamline at Fermilab was in service for about 3.5 years until neutrino yield degradation was observed during the second half of its life. The probable causes for the target performance degradation were partially attributed to radiation damage and possibly cracking caused by reduction in thermal shock resistance. The target was 95 cm long and consisted of 48 segmented POCO ZXF-5Q grade graphite fins. The target was exposed to a total of 6.1 x $10^{20}$ protons (120 GeV) over the its lifetime, amounting to a peak accumulated DPA of about 0.6.

During recovery of the graphite fins from the target, several fins were discovered to be cracked near the centerline of beam passage, with some of these fins separated from the cooling water tubes. At the time, it was unclear whether separation and fracture of the fins occurred during operation or during the disassembly process, and therefore PIE studies were initiated at Pacific Northwest National Laboratory to explore the fins [5].

Bulk dimensional measurements on recovered fins were first carried out, revealing up to 4% swelling across the middle of the more upstream fin. These results are consistent with the higher proton fluence in the middle (along beam path) as well as the greater damage rate at the upstream end of the target.

Elemental analysis on the surface of fractured fins suggested that loose particulate and impurities might have migrated to the interior of the cracked fins during operation. Furthermore, optical and electron microscopy showed propagation ripple patterns on the fracture surface that supports the notion that the crack initiated at the center of the fins and during operation.

On the other hand, TEM analyses did not reveal any noticeable signs of displacement damage near the beam spot regions, possibly due to the lower irradiation temperature of the fins during operation (50-200 °C). Future work on the fins at the University of Oxford will include micro-cantilever and hardness measurements (nano-indentation) to extract mechanical properties of the irradiated graphite fins to help quantify any radiation damage effects.

### 2.3. NuMI Beryllium Primary Beam Window Study

An irradiated PF-60 grade beryllium thin disc (0.25 mm thick) was recovered from the Fermilab NuMI primary beam window assembly and sent to the University of Oxford for PIE activities [6]. Over its lifetime, the Be disc was exposed to a total of 1.6 x $10^{21}$ protons (120 GeV), equivalent to about 0.5 peak DPA, with an irradiation temperature of about 50 °C. Cracks had appeared near the central beam region on the beryllium disc during the removal process from the window assembly.

PIE analyses showed clear indication of crack morphology shifts at high doses from transgranular to grain boundary fracture indicating hardening of the crystal matrix within grains and/or grain boundary weakening, shown in Figure 3.

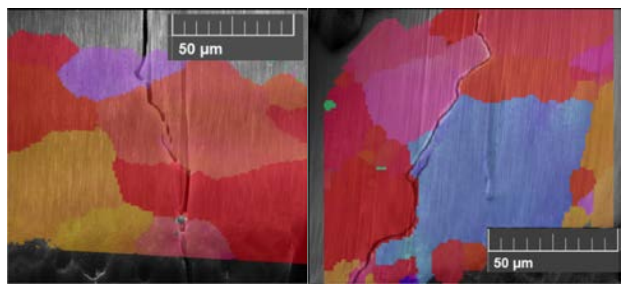

Figure 3. Electron Back-Scatter Diffraction (EBSD) images showing cracks in NuMI Be beam window irradiated to 0.29 DPA (left, transgranular fracture), and 0.44 DPA (right, intergranular fracture).

Atom Probe Tomography (APT) studies were also carried out on the Be disc and revealed Lithium production through the transmutation process, distributed homogeneously through the material. Measured Li concentration matched MARS [7] numerical simulation results to within 25%.

Ongoing and future work on the NuMI Be beam window will involve micro-mechanical testing using micro-cantilevers and nano-indentation to extract mechanical properties. Preliminary results have already indicated significant hardening and increase in elastic modulus for the irradiated material.

### 2.4. Helium Implantation of Be Study

In order to replicate the radiation damage effects in beryllium due to high energy protons, low energy helium ions were implanted into beryllium specimens at the University of Surrey Ion Beam Center. Irradiation was conducted with 2 MeV Helium ions, at temperatures of 50 °C and 200 °C and up to 0.1 DPA and 2000 atomic ppm He. Hardness measurements were then carried out on the beryllium specimens at the University of Oxford [8].

As shown in Figure 4, nano-indentation measurements revealed significant hardening at 0.1 DPA which was primarily due to implanted helium content rather than displacement damage. TEM analyses confirmed evidence of the He implantation by revealing nanometer scale black dots. Ongoing work consists of micro-cantilever measurements to extract mechanical properties, as well as plans for future low energy irradiations at higher doses and temperatures.

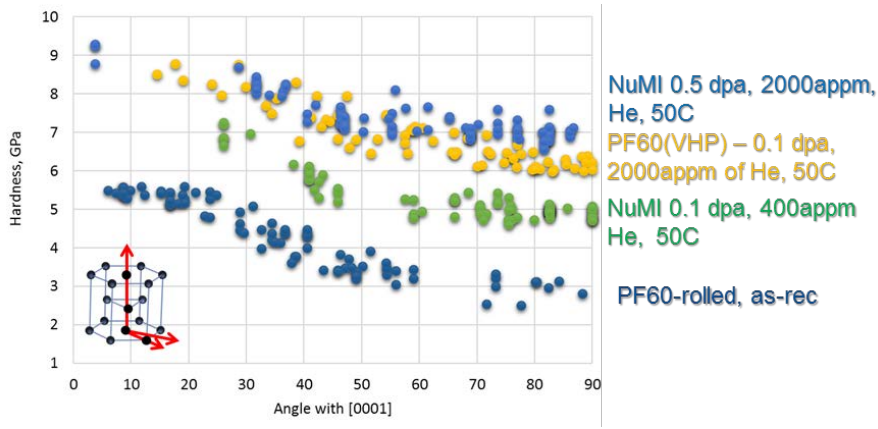

Figure 4. Hardening in He implanted beryllium compared to as-received PF-60 beryllium.

### 2.5. In-Beam Thermal Shock of Be Study

To investigate the thermal shock response of beryllium, several grades of Be specimens of varying thicknesses were exposed to high energy high intensity beams at CERN's HiRadMat facility [9]. Peak beam pulse intensities of up to $2.8 \times 10^{13}$ protons per 7.2 μs pulses with beam spot sizes less than 0.3 mm (Gaussian sigma radius) were imposed on Be specimens to push them to their failure limit through plastic deformation at high temperatures [10].

PIE work was carried out at the University of Oxford and Figure 5 shows profilometry measurements of the out-of-plane plastic deformation induced in different 0.75 mm thick Be grade specimens from increasing beam intensities. Preliminary results indicate that the S200FH grade generally showed the least amount of plastic deformation, while the S200F grade showed the largest amount of plastic deformation.

Multiple beam pulses at the same location on the Be discs in Array 3 also demonstrated plastic deformation ratcheting.

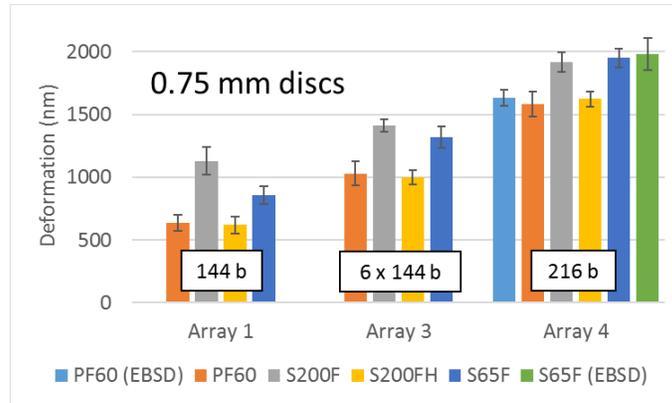

Figure 5. Profilometry measurements of out-of-plane plastic deformation induced by high intensity proton beam at HiRadMat.

A recently developed S200FH beryllium Johnson-Cook strength model was developed to compare Finite Element Method (FEM) simulations with experimental measurements. As shown in Figure 6, very good agreement to within 10% was obtained between FEM simulation results and measurements for the S200FH case at the highest beam intensity. Note that the other Be grades have different yield strengths and chemical compositions and therefore incurred different amounts of out-of-plane deformation. Data analysis and further validation of the Johnson-Cook model is ongoing. A follow-up experiment at HiRadMat in 2018 is being currently being designed to explore deformation at even higher beam intensity as well as compare the performance of irradiated versus non-irradiated materials, including beryllium, graphite, and titanium alloys.

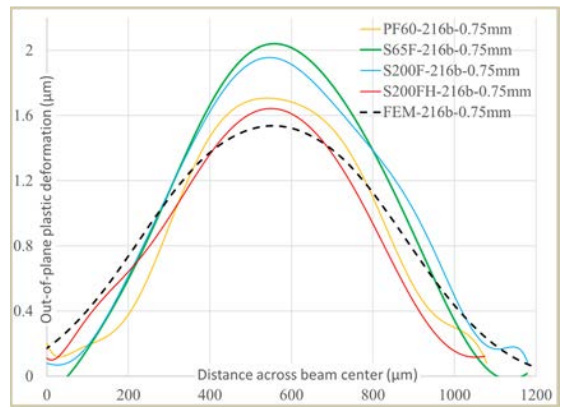

Figure 6. Johnson-Cook (Be S200FH) simulation of out-of-plane plastic deformation (dashes) compared to profilometry measurements.

## 3. FUTURE RADIATE ACTIVITIES

### 3.1. Multi-Material BLIP Irradiation and PIE

A material irradiation campaign utilizing the BNL BLIP facility with 181 MeV protons was started in early 2017, with almost all RaDIATE participating institutions playing a significant role. The materials included in the irradiation run were relevant to each of the participants' institution and future programs, and consisted of beryllium, graphite, aluminum alloys, titanium alloys (including 3D printed material), silicon, silicon carbide, TZM, CuCrZr and iridium [11]. The specimen capsules, like the one shown in Figure 1, already received partial irradiation in 2017, and is expected to complete the 8-week irradiation campaign in early 2018 when isotope production resumes at BLIP. New materials such as Ta2.5W, Mo-coated CfC and MoGr, and new grades of Ti will be added to the second irradiation phase. After the irradiation and sufficient cool-down time, hundreds of individual irradiated test specimens will be available for PIE work, which is expected to be performed later in 2018 and 2019.

### 3.2. HiRadMat in-beam thermal shock test

As mentioned earlier, a follow-up thermal shock experiment at CERN's HiRadMat facility is currently being designed to take place in 2018. The new experiment will expose non-irradiated and irradiated specimens (from the BLIP irradiation) of beryllium, graphite, titanium alloy, glassy carbon and silicon to the high intensity 440 GeV proton beam to evaluate their thermal shock response during PIE measurements. Some specimens will be partially instrumented to record real-time temperature and strain induced by beam pulses. This will allow a direct comparison of irradiated versus non-irradiated material, validation of simulation data and techniques, and prediction of how highly irradiated material responds to the unique loading environment of intense proton beam. Other novel materials such as electro-spun nanofiber mats and foam materials will also be included in the test matrix.

### 4. CONCLUSIONS

As briefly covered in this paper, the RaDIATE collaboration is growing and has initiated several radiation damage and thermal shock studies over the last five years, with many participants actively involved. Some accelerator facilities are currently limited in beam power due to target survivability issues, and future multi-MW accelerator upgrades and facilities will pose even greater challenges in order to safely and reliably operate these facilities while maximizing the physics benefits. Therefore, materials R&D activities by the global accelerator targets community within the framework of the RaDIATE collaboration are moving forward at a fast pace to help meet the future challenges. The primary objective of the collaboration is to harness existing expertise in nuclear materials and accelerator targets to generate new and useful materials data for application within the accelerator targetry community.

### ACKNOWLEDGMENTS

This manuscript has been authored by Fermi Research Alliance, LLC under Contract No. DE-AC02-07CH11359 with the U.S. Department of Energy, Office of Science, Office of High Energy Physics.

Submitted on behalf of the RaDIATE collaboration. Participating institutions include Fermi National Accelerator Laboratory (FNAL), Science and Technology Facilities Council (STFC), the Chancellor Masters and Scholars of Oxford University (Oxford), Brookhaven National Laboratory (BNL), Pacific Northwest National Laboratory (PNNL), Oak Ridge National Laboratory (ORNL), Michigan State University (MSU), European Spallation Source (ESS), Los Alamos National Laboratory (LANL), Argonne National Laboratory (ANL), Centro de

Investigaciones Energéticas, Medioambientales y Tecnológicas (CIEMAT), European Organization for Nuclear Research (CERN), High Energy Accelerator Research Organization (KEK), and Japan Atomic Energy Agency (JAEA).